\documentclass[12pt,preprint]{aastex}

\newcommand{\etal }{{et al.} }
\newcommand{\msun}{\thinspace M_\odot} 
 
\newcommand{\rsun}{\thinspace R_\odot} 
\newcommand{\vect}[1]{\mbox{\boldmath$#1$}}
\def\lesssim{\mathrel{\hbox{\rlap{\hbox{\lower4pt\hbox{$\sim$}}}\hbox{$<$}}}}
\def\gtrsim{\mathrel{\hbox{\rlap{\hbox{\lower4pt\hbox{$\sim$}}}\hbox{$>$}}}}
\newcommand{\cm}{\,{\rm cm}^{-3} } 
 
\newcommand{\nc}{n_{\rm c} } 
\newcommand{\rcri}{R_{\rm c} }

\newcommand{\tc}{t_{\rm c}}

\newcommand{\dfrac}[2]{{\displaystyle \frac{#1}{#2}} }

\shorttitle{The Origin of Circumstellar Disk}
\shortauthors{Machida \& Matsumoto 2010}

\begin{document}
\title{The Origin and Formation of the Circumstellar Disk}
\author{Masahiro N. Machida\altaffilmark{1} and Tomoaki Matsumoto\altaffilmark{2}} 
\altaffiltext{1}{National Astronomical Observatory of Japan, Mitaka, Tokyo 181-8588, Japan; masahiro.machida@nao.ac.jp}
\altaffiltext{2}{Faculty of Humanity and Environment, Hosei University, Fujimi, Chiyoda-ku, Tokyo 102-8160, Japan; matsu@hosei.ac.jp}

\begin{abstract}
The formation and evolution of the circumstellar disk in the collapsing molecular cloud is investigated from the prestellar stage resolving both the molecular cloud core and the protostar itself. 
In the collapsing cloud core, the first (adiabatic) core appears prior to the protostar formation.
Reflecting the thermodynamics of the collapsing gas, the first core is much more massive than the protostar.
When the molecular cloud has no angular momentum, the first core falls onto the protostar and disappears several years after the protostar formation.
On the other hand, when the molecular cloud has an angular momentum, the first core does not disappear even after the protostar formation, and directly evolves into the circumstellar disk with a Keplerian rotation.
There are two paths for the formation of the circumstellar disk.
When the initial cloud has a considerably small rotational energy, two nested disks appear just after the protostar formation.
During the early main accretion phase, the inner disk increases its size and merges with the outer disk (i.e. first core) to form a single circumstellar disk with a Keplerian rotation.
On the other hand, when the molecular cloud has a rotational energy comparable to observations, a single Keplerian disk that corresponds to the first core  already exists prior to the protostar formation.
In such a cloud, the first core density gradually increases, maintaining the Keplerian rotation and forms the protostar inside it.
Thus, the protostar is born in the Keplerian disk.
In other words, a massive disk already exists before the protostar formation.
In each case, the protostar at its formation is already surrounded by a massive circumstellar disk.
The circumstellar disk is about 10-100 times more massive than the protostar in the main accretion disk.
Such disks are favourable sites for the formation of binary companions and gas-giant planets.

\end{abstract}
\keywords{accretion, accretion disks: ISM: clouds---stars: formation---stars: low-mass, brown dwarfs: planetary systems: protoplanetary disks}

\section{Introduction}
\label{sec:intro}
A star is born with a circumstellar disk in the molecular cloud core.
Molecular clouds have an angular momentum \citep{arquilla86,goodman93,caselli02} that can form a disk in the star-formation process.
Thus, disk formation is a natural consequence of the conservation of angular momentum in the collapsing cloud core.
Observations have also supported the existence of a circumstellar disk around young stellar objects.
The circumstellar disks have been identified by spectral energy distributions, and are resolved in their scattered light \citep[see review of ][]{watson07}.
Thus, there is no room for doubt about the emergence and presence of circumstellar disks in the star-formation process.

The circumstellar disk plays important roles in star and planet formation.
The planets are formed in the circumstellar (or protoplanetary) disks.
Recent theoretical studies state that planets are formed according to the gravitational instability scenario with massive disks and according to the core accretion scenario with less massive disks \citep{durisen07}.
Thus, the mass of the circumstellar disk determines the planet-formation mode and properties of planets.
In addition, the protostellar outflows are driven by the circumstellar disk \citep{pudritz07}.
It is considered that these flows are closely related to the angular momentum transfer, and they determine the star-formation efficiency \citep{matzner00}. 
Therefore, it is necessary to investigate the formation and properties of the circumstellar disk to understand both star and planet formation.
Nevertheless, so far, the formation process of the circumstellar disk in the collapsing cloud core has rarely been investigated.

In a classical star-formation scenario, it is envisaged that the circumstellar disk begins to form after the protostar formation and increases its size with time.
This picture is evoked by the idea that a small part of the natal cloud having the lowest angular momentum first collapses and forms the protostar,  while the remainder of the cloud, which has larger angular momentum collapses later, and forms the circumstellar disk.
However, in this picture, the thermal evolution of collapsing (or infalling) gas is not taken into account.
Recent star formation scenarios including thermal evolution,  have shown that the transient object that is called  the first core appears prior to the protostar formation  \citep{masunaga00}.
In the collapsing cloud core, the gas collapses isothermally for $n \lesssim 10^{10}\cm$, while it collapses adiabatically for $n \gtrsim 10^{10}\cm$ and forms the first (adiabatic) core at $n\simeq10^{10}-10^{12}\cm$.  
Although the first core is in a nearly hydrostatic equilibrium state, its central density gradually increases because the first core increases its mass by the gas accretion \citep{masunaga98}.
When the central density exceeds $n\gtrsim10^{16}\cm$ (or temperature $T \gtrsim 2000$\,K), the molecular hydrogen begins to dissociate and the central region rapidly collapses again (the so-called second collapse).
Finally, the collapsing gas becomes adiabatic again and the protostar forms at $n\gtrsim10^{20}\cm$ \citep{larson69,masunaga00}.
Because the second collapse occurs only in  a small part of the first core, the first core does not disappear just after the protostar formation.
In a spherically symmetric calculation, the first core falls onto the central protostar and disappears several years after the protostar formation \citep{masunaga00}.
However, the rotational effect is not included in the spherically symmetric calculation.
In reality, because the first core is supported by not only the thermal pressure but also the rotation, it does not disappear in such a short duration without effective  angular momentum transfer \citep{saigo06}.

Recently, \citet{machida10} and \citet{bate10}  pointed out that the first core is the origin of the circumstellar disk \citep[see also][]{bate98}, indicating that the disk appears before the protostar formation and is more massive than the protostar at the protostar formation epoch \citep{inutsuka10}.
Such massive disks tend to exhibit the gravitational instability that contributes to the angular momentum transfer through the nonaxisymetric gravitational torque, and binary or planet formation  through fragmentation.
To investigate the formation and evolution of the circumstellar disk in the collapsing cloud core, we need a three-dimensional simulation that covers both the circumstellar disk and its natal cloud with adequate spatial resolution.
However, because such a calculation requires a considerably higher spatial resolution, only a few studies have investigated the formation of the circumstellar disk from the prestellar stage.
In many other calculations, the evolution of circumstellar disks after the main accretion phase was investigated with artificial settings in which the disk properties are artificially modeled \citep[see review of ][]{durisen07}.
However, the evolution of the circumstellar disk even after the main accretion phase should be investigated from the prestellar stage to give adequate initial condition. 
Otherwise, artificial initial settings may determine the final fate of the circumstellar disk.

\citet{walch09a,walch09b} and \citet{machida10} calculated the formation of the circumstellar disk from the prestellar stage (i.e. from the starless molecular cloud core stage) up to the main accretion stage in an unmagnetized cloud.
However, they did not resolve the protostar itself to perform a long-term calculation of the circumstellar disk.
\citet{bate98} investigated the protostar formation with a sufficient spatial resolution up to the protostar formation.
The cloud evolution in both phases (before and after the protostar formation) resolving the protostellar radius in an unmagnetized cloud was only calculated by \citet{bate10}. 
However, that study did not explore the evolution of the circumstellar disk just after the protostar formation.

In summary, so far, the evolution of the circumstellar disk in the proximity of the protostar just after the protostar formation has not been investigated with a multidimensional simulation resolving the protostar (or protostellar size).
If the first core that is formed before the protostar formation falls onto the protostar and disappears for a short duration, the circumstellar disk gradually forms after the protostar formation, as envisaged in the classical picture.
Instead, as pointed out by \citet{bate10} and \citet{machida10}, if the first core remains longer, it becomes the circumstellar disk that is considerably more massive than the protostar.
In this study, we calculate the cloud evolution from the prestellar stage up to the period after the protostar formation resolving the protostellar radius of $<0.01$\,AU  and investigate the formation process of the circumstellar disk in the proximity of the protostar.
The structure of this paper is as follows. 
The framework of our models and the numerical method are given in \S 2. 
The numerical results are presented in \S 3. 
We discuss the further evolution of the disk in \S 4, and summarize our results in \S 5.

\section{Model Settings}
\label{sec:model}
To study the evolution of rotating gas clouds and the formation of circumstellar disks, we solve the equations of hydrodynamics including self-gravity:
\begin{eqnarray} 
& \dfrac{\partial \rho}{\partial t}  + \nabla \cdot (\rho \vect{v}) = 0, & \\
& \rho \dfrac{\partial \vect{v}}{\partial t} 
    + \rho(\vect{v} \cdot \nabla)\vect{v} =
    - \nabla P -      \rho \nabla \phi, & 
\label{eq:eom} \\ 
& \nabla^2 \phi = 4 \pi G \rho, &
\end{eqnarray}
where $\rho$, $\vect{v}$, $P$, and $\phi$ denote the density, velocity, pressure and gravitational potential, respectively. 
To mimic the temperature evolution calculated by \citet{masunaga00}, we adopt the piece-wise polytropic equation  (see, eq~[5] of \citealt{machida07}).
In reality, the thermal evolution in the collapsing cloud may have to be investigated with radiation hydrodynamics code. 
However, \citet{bate10} investigated the evolution of the collapsing cloud using radiation smoothed particle hydrodynamics code, and commented that the cloud evolution with radiation hydrodynamics is almost identical to the evolution obtained from the calculation with the barotropic equation of state before the protostar formation \citep[see also,][]{tomida10}.
In this study, we investigate the cloud evolution with the barotropic equation of state up to the epoch just after the protostar formation.
We think that, in this period, the cloud evolution and formation of the circumstellar disk can be safely investigated by using the barotropic approximation.
This treatment reduces CPU time and makes it possible to study cloud evolution with sufficiently high spatial resolution.

As the initial state, we take a spherical cloud with critical Bonnor--Ebert (BE) density profile, in which a uniform density is adopted outside the sphere ($r > \rcri$).
For the BE density profile, we adopt the central density of $\nc =  10^{6}\cm$ and isothermal temperature of $T=10$\,K. 
For these parameters, the critical BE radius is $\rcri = 4.8\times10^3$\,AU.
To promote the contraction, we increase the density by a factor of $f$=1.68, where $f$ is the density enhancement factor that represents the stability of the initial cloud.
With $f=1.68$, the initial cloud has (negative) gravitational energy twice that of thermal energy.
The mass within $r < \rcri$ is $M = 0.8\msun$.
Initially, the cloud rotates rigidly with an angular velocity $\Omega_0$ around the $z$-axis.
We parameterized the ratio of the rotational to the gravitational energy ($\beta_0 \equiv E_r/E_g$) inside the initial cloud.
We calculated four different models with different $\beta_0$.
Model names, initial angular velocities $\Omega_0$, and $\beta_0$ are summarized in Table~1.

The time step for the calculation becomes increasingly short as the cloud collapses (or the central density increases).
When the time step becomes extremely short, we cannot calculate the evolution of the cloud and circumstellar disk for long durations. 
To avoid this, we adopt a sink at the centre of the cloud after the protostar formation.
To model the protostar, we adopt a sink around the center of the computational domain when the number density $n$ exceeds $ > 3\times10^{21}\cm$ at the cloud centre.
In the region $r < r_{\rm sink} = 2\,\rsun$, the gas having a number density of $n > 3\times 10^{21}\cm$ is removed from the computational domain and added to the protostar as gravity in each timestep \citep[for details, see][]{machida09a,machida10}.
\citet{masunaga00} showed that the protostar forms at $n\sim10^{20}-10^{21}\cm$ with a radius of $\sim2\rsun$ in the collapsing cloud.
This treatment of the sink makes it possible to calculate the evolution of the collapsing cloud and circumstellar disk for a longer duration.
However, in our calculations, the central density never exceeds $n>3\times10^{21}\cm$ except for the model with $\beta_0=0$ (i.e., non-rotating model).
Thus, we calculated the cloud evolution up to  the end of the calculation without a sink for the model with $\beta_0 \ne 0$, while with a sink for model with $\beta_0=0$.
In each model, we stop the calculation about 4\,yr after the protostar formation.

To calculate over a large spatial scale, the nested grid method is adopted \citep{machida05a,machida06a}.
Each level of a rectangular grid has the same number of cells, $  128 \times 128 \times 16 $.
The calculation is first performed with five grid levels ($l=1$--$5$).
The box size of the coarsest grid $l=1$ is chosen to be $2^5 \rcri$.
Thus, a grid of $l=1$ has a box size of $1.5\times 10^5$\,AU.
A new finer grid is generated before the Jeans condition is violated.
The maximum level of grids is restricted to $l_{\rm max} \leqq 21$.
The $l=21$ grid has a box size of 0.14\,AU and a cell width of $1.1\times10^{-3}$\,AU.
With this method, we cover eight orders of magnitude in spatial scale.
In other words, we can spatially cover both the molecular cloud core and protostar itself.

\section{Results}
\label{sec:results}
The disk-formation process depends on the rotation rate of the initial cloud, because the cloud rotation makes the circumstellar disk in the star-formation process.
In this study, we calculated the formation of the protostar and circumstellar disk in the molecular cloud core for four different models in which the rotational energy of the initial cloud is parameterized in the range of $0<\beta_0<10^{-3}$. 
The observations have shown that the molecular clouds have rotational energy in the range of $10^{-4}\lesssim \beta_0 \lesssim 0.07$ with an average of $\beta_0 = 0.02$ \citep{goodman93,caselli02}.
Recent numerical studies showed that the cloud with rapid rotation $\beta_0\gtrsim10^{-2}$ tends to show fragmentation or binary formation before the protostar formation \citep{matsu03}.
However, it is difficult to investigate the formation process of the circumstellar disk in a binary system.
Thus, in this study, we limited the rotational parameter in the range of $\beta_0 \le 10^{-3}$ to focus on the circumstellar disk around the single protostar. 
In this section, first, we show the cloud evolution for the model with a relatively slow rotation rate ($\beta_0\le10^{-4}$).
Then, the evolution of a relatively rapidly rotating cloud ($\beta_0=10^{-3}$) is described.
The evolution of the cloud with larger rotational energy ($\beta_0>10^{-3}$) is discussed in \S\ref{sec:frag}.

\subsection{Evolution of Cloud with a Relatively Slow Rotation}
\label{sec:mod}
\begin{figure}
\includegraphics[width=150mm]{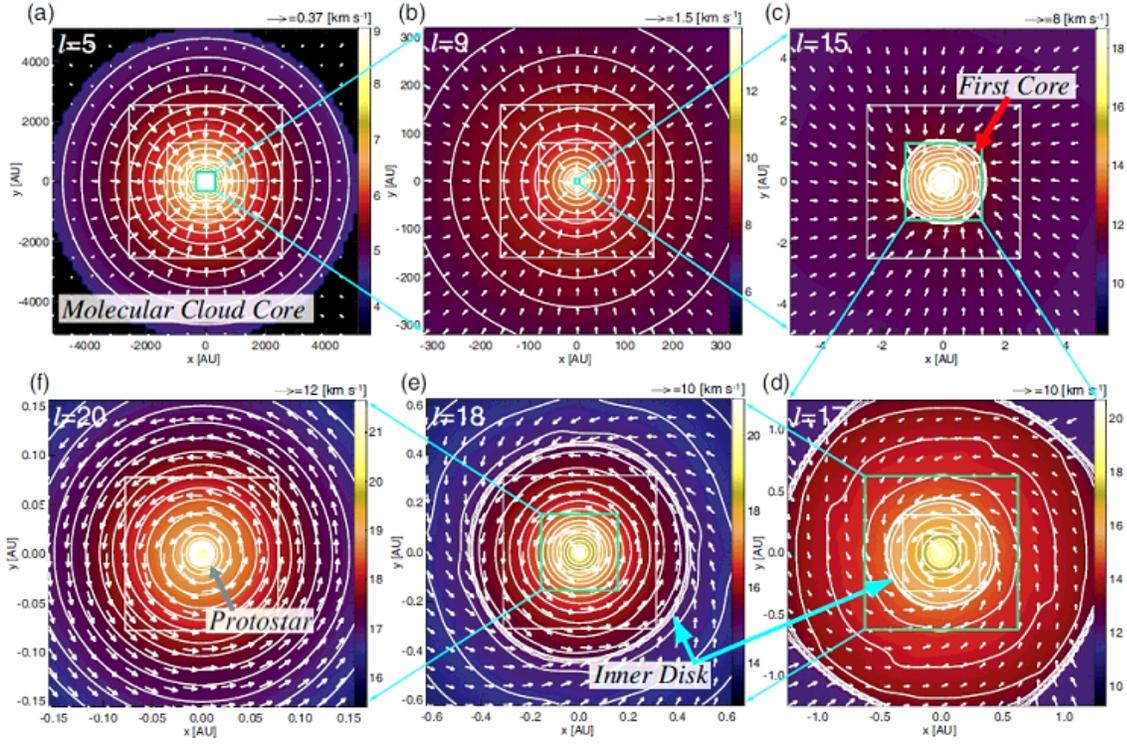}
\caption{
The density distribution ({\it colour and contours}) and velocity vectors ({\it arrows}) on the equatorial plane at the epoch just after the protostar formation ($t=5.064\times 10^4$\,yr, $\tc=1.2083$\,yr) for model 2 ($\beta_0=10^{-4}$) at various spatial scales.
Each object (molecular cloud core, first core, circumstellar disk and protostar) is identified  by  an arrow.
}
\label{fig:1}
\end{figure}

\begin{figure}
\includegraphics[width=150mm]{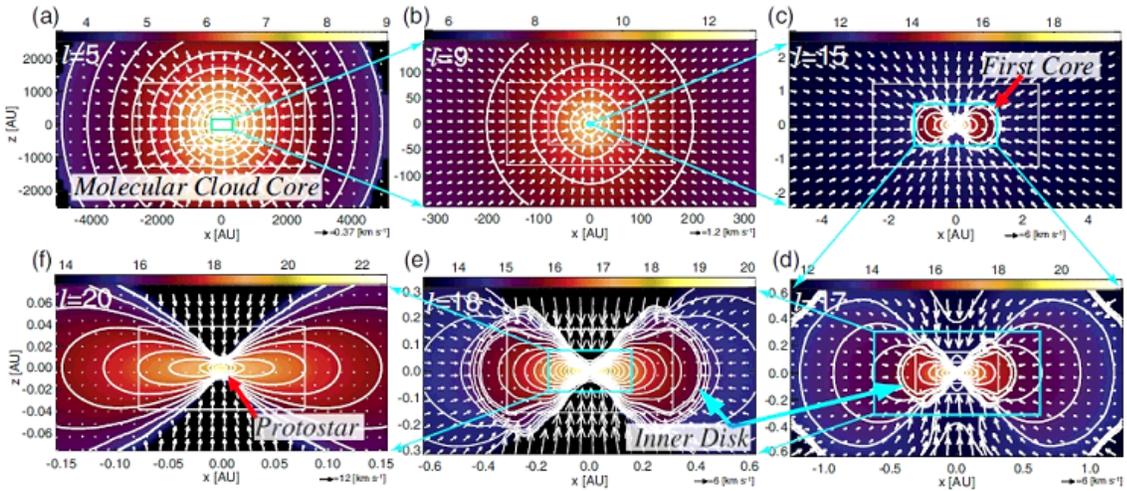}
\caption{
Same as Fig.~\ref{fig:1} but on the $y=0$ plane.
}
\label{fig:2}
\end{figure}

Figures~\ref{fig:1} and \ref{fig:2} show the density distribution and velocity vectors on the $z=0$ (Fig.~\ref{fig:1}) and $y=0$ (Fig.~\ref{fig:2}) planes for a model with $\beta_0=10^{-4}$ (model 2) with different spatial scales  $\tc=1.2083$\,yr after the protostar formation.
In these figures, Figures~\ref{fig:1}{\it a} and \ref{fig:2}{\it a} cover the entire region of the molecular cloud core with a box size of $\sim8000$\,AU, while Figures~\ref{fig:1}{\it f} and  \ref{fig:2}{\it f} cover the protostar and a part of the disk with a box size of $\sim0.3$\,AU.
For this model, the protostar forms in the collapsing cloud $t=5.064\times10^4$\,yr after the calculation begins (i.e. after the initial cloud begins to collapse).
In this paper, we define $\tc$ as the elapsed time after the protostar formation where the protostar formation epoch ($\tc=0$) is defined as the epoch when the central density reaches $\nc=10^{20}\cm$ in the collapsing cloud core.
In addition, we define $t$ as the elapsed time after the calculation begins (or the initial cloud begins to collapse).

Figures~\ref{fig:1}{\it a} and \ref{fig:2}{\it a} show that a molecular cloud with a size of $\sim 8000$\,AU maintains a spherical structure even after the protostar formation, because the gas begins to collapse in the small area of the centre of the cloud and the outer envelope maintains its nearly initial structure, as shown in \citet{larson69}.
As seen in Figures~\ref{fig:1}{\it b} and \ref{fig:2}{\it b}, the gas collapses spherically towards the centre of the cloud in a large scale of $\gg 1$\,AU, because the rotation hardly affects the dynamical evolution of the cloud at such a scale \citep{machida05a}.
In the collapsing cloud, the gas becomes optically thick against the dust cooling and the first core surrounded by the shock appears when the number density exceeds $n \gtrsim 10^{10}\cm$ \citep{masunaga00}.
When the cloud has an angular momentum, the first core has a disk-like structure.
The disk-like structure surrounded by the shock in Figures~\ref{fig:1}{\it c} and \ref{fig:2}{\it c} corresponds to the first core.
Figures~\ref{fig:1}{\it b-d} show that the gas falls radially towards the first core outside the first core, while the gas rotates rapidly inside the first core.
This is because, inside the first core,  the gas collapses very slowly in the radial direction and the rotation  dominates the radial motion after the gas becomes adiabatic (i.e. after the first core formation).

Inside the first core, there is another structure with a radius of $\sim0.4$\,AU surrounded by the disk-like shock, as seen in panels {\it d} and {\it e} of Figures~\ref{fig:1} and \ref{fig:2}.
This disk-like object is formed inside the first core after the protostar formation.
In this paper, we call this object (i.e., the disk around the protostar) the inner disk.
In addition, the protostar with a radius of $\sim0.02$\,AU exists with a nearly spherical structure inside the inner disk, as seen in Figures~\ref{fig:1}{\it f} and \ref{fig:2}{\it f}.
Thus, after the protostar formation, three different structures appear inside the collapsing molecular cloud core: two nested disks (disk-like first core and inner disk) and protostar.

After the first core formation,  further rapid collapse is induced in a small central part of the first core owing to the dissociation of molecule hydrogen when the number density exceeds $n\gtrsim10^{16}\cm$ \citep{larson69,masunaga00}.
In such a collapsing region, the gas becomes adiabatic again after the dissociation of molecule hydrogen is completed and a protostar with a shock (or the second adiabatic core) appears in the region of $n\gtrsim10^{20}\cm$.
In this model, then, the inner disk appears around the protostar after the protostar formation, because the first core has an angular momentum and the rotation of the infalling gas from the first core cannot be neglected as it approaches the protostar.
Therefore, the disk structure is formed inside the first core because of the rotation.
Note that the first core appears in the collapsing cloud even without rotation, because the first core is supported by not only the rotation but also thermal pressure, while no inner disk appears without rotation.
Figures~\ref{fig:1} and \ref{fig:2} clearly show two nested disks inside the molecular cloud core outside the protostar.

\begin{figure}
\includegraphics[width=150mm]{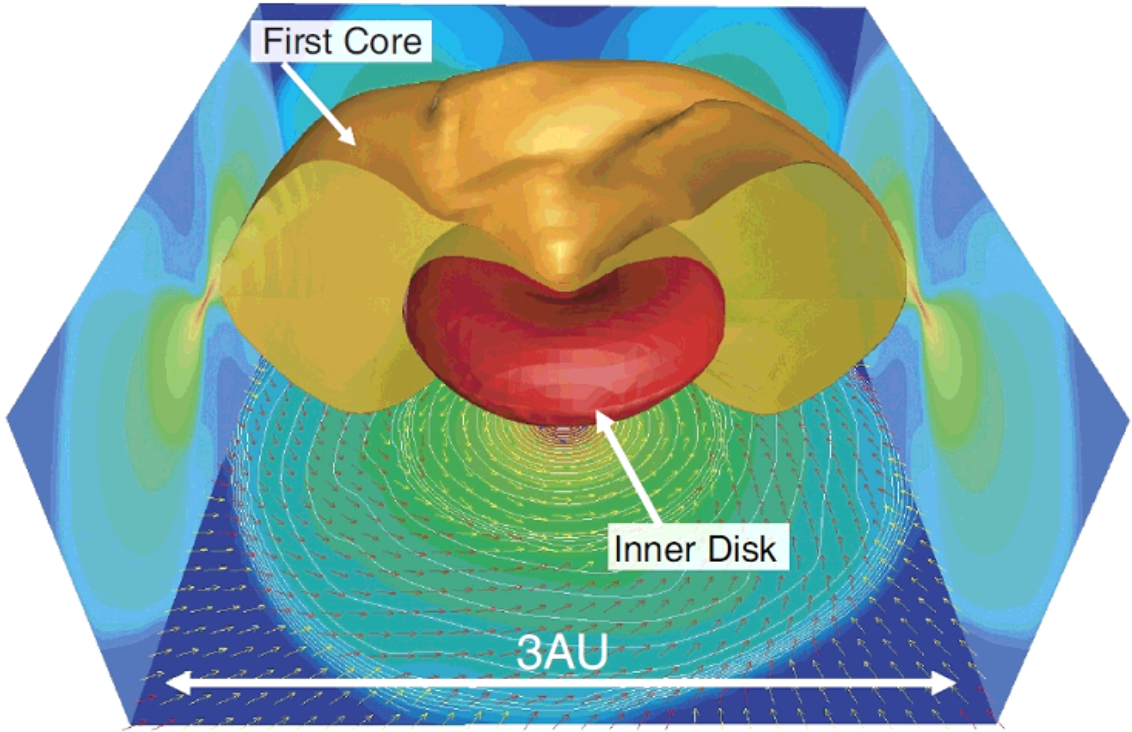}
\caption{First core ({orange isodensity surface}) and inner disk ({\it red isodensity surface}) for model 2 ($\beta_0=10^{-4}$) at $\tc=1.664$\,yr after the protostar formation  are plotted in three dimensions.
The density distribution on the $x=0$, $y=0$ and $z=0$ plane is projected onto each wall surface.
The velocity vectors on the $z=0$ plane are also projected onto the bottom wall surface.
}
\label{fig:3}
\end{figure}

The nested disks can also be seen in Figure~\ref{fig:3}, in which the first core and inner disk are plotted by the orange and red iso-density surfaces, respectively.
This figure clearly shows that the thin inner disk is surrounded by the thick torus-like disk (i.e., the first core).
In the figure, the velocity vectors on the equatorial plane are projected onto the bottom wall surface.  
The direction of the velocity vector is suddenly changed at the surface of the first core;  the gas  falls radially towards the centre of the cloud outside the first core, while it falls slowly with rapid rotation inside the first core.
In addition, inside the inner disk (red iso-density surface), the azimuthal greatly velocity dominates the radial velocity (i.e. $v_\phi \gg v_r$). 
Thus, on the equatorial plane, the gas inside the inner disk is rotating rapidly without infalling towards the protostar.

\begin{figure}
\includegraphics[width=150mm]{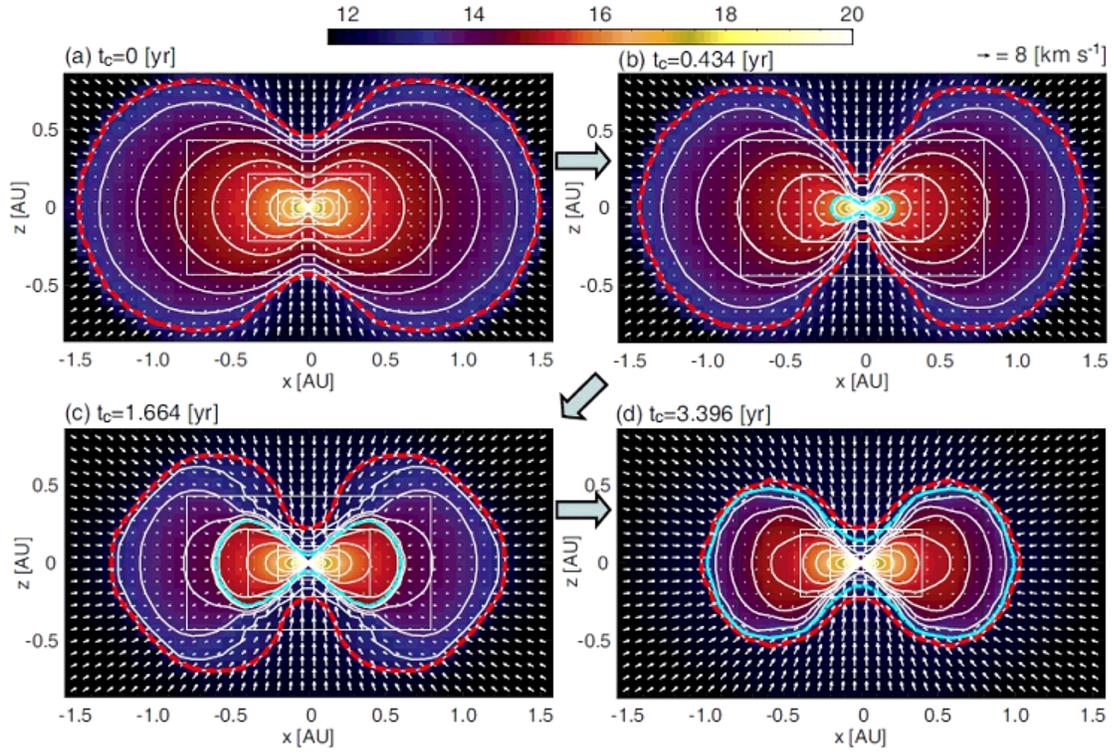}
\caption{
Time sequence for model 2 ($\beta_0=10^{-4}$).
The density distribution ({\it colour and contours}) and velocity vectors ({\it arrows}) on the $y=0$ plane are plotted for various epochs.
In each panel, the inner shock that corresponds to the circumstellar disk is plotted by a blue dotted line, while the outer shock that corresponds to the remnant first core is plotted by a red dotted line.
}
\label{fig:4}
\end{figure}

Figure~\ref{fig:4} shows the time sequence of the first core and inner disk after the protostar formation, in which each panel has a same box size of $\sim3$\,AU.
In the figure, the outer shock surface corresponding to the first core is plotted by the broken red line, while the inner shock surface corresponding to the  inner disk is plotted by the broken blue line.
The first core has an ellipsoidal structure at the protostar formation epoch (Fig.~\ref{fig:4}{\it a}).
Then, after the inner disk appears around the protostar (or after the protostar formation), the first core thins in the region just outside the inner disk.
In Figure~\ref{fig:4}{\it b}, we can see the sharp drop of density distribution in the vertical direction in the range of $\vert x \vert < 0.5\,{\rm AU}$.
The thermal energy decreases in the centre of the first core, where the equation of state is changed from $\gamma=1.4$ to $\gamma=1.1$ because of the dissociation of molecular hydrogen. 
Thus, the gas inside the first core (or in the central region of the first core)  falls vertically onto the protostar or the inner disk, because the vertical direction is mainly supported by the thermal pressure gradient force not by the centrifugal force.
After the protostar formation, the inner disk increases its size, while the first core slightly decreases its size, as seen in Figures~\ref{fig:4}{\it b}-{\it c}.
This is because the region inside the first core gradually collapses to accrete onto the inner disk.
In addition, the first core increases its mass by the gas accretion and shrinks its size \citep{saigo06}.
Finally, two shocks  (or two disks) composed of the shrinking first core and expanding inner disk merge to form  a single disk (or single shock) $\tc=3.4$\,yr after the protostar formation, as shown in Figure~\ref{fig:4}{\it d}.
After the merger, the single disk increases its size with time.

To confirm this merger, the radial distribution of the density ({\it a}), radial velocity ({\it b}),  azimuthal and Kepler velocities ({\it c}) and ratio of the radial velocity to the azimuthal velocity ({\it d}) are plotted in Figure~\ref{fig:5}, in which each quantity is azimuthally averaged.
At the protostar formation epoch ($\tc=0$\,yr; black line),  the density profile shows only a single shock at $r\sim1.7$\,AU (Fig.~\ref{fig:5}{\it a}).
At the shock surface that corresponds to the surface of the first core, the radial velocity suddenly increases to approach  zero (Fig.~\ref{fig:5}{\it b}).
Inside the first core ($r<1.7$\,AU), the radial velocity gradually decreases again in the range of $0.02\,{\rm AU}\lesssim r \lesssim 1.7\,{\rm AU}$, while it increases in the range of $r\lesssim 0.02$\,AU.
Because the protostar ($n\gtrsim10^{20}\cm$) has a size of $\sim0.01$\,AU, the infall (or negative radial) velocity approaches  zero near the protostellar surface.
The azimuthal velocity suddenly increases at the surface of the first core (Fig.~\ref{fig:5}{\it c}).
Figure~\ref{fig:5}{\it d} shows that the radial velocity greatly dominates the azimuthal velocity outside the first core, while the azimuthal velocity dominates the radial velocity inside the first core.
This indicates that the first core is rotating rapidly and collapses slowly.

The blue and red lines in Figure~\ref{fig:5}{\it a} show two nested shocks in the range of $0.1\, {\rm AU} \lesssim r \lesssim 2\, {\rm AU}$: the outer shock corresponding to the first core is located at $r\sim1-2$\,AU, while the inner shock corresponding to the inner disk is located at $r\sim0.1-0.5$\,AU.
The figure indicates that the inner shock is outwardly expanding, while the outer shock is inwardly shrinking.
In Figure~\ref{fig:5}{\it b}, we can clearly see two shocks for red and blue lines, in which the radial velocity is considerably small except for the very proximity of the protostar ($r<0.01$\,AU) inside the inner shock (or the inner disk).
Figure~\ref{fig:5}{\it c} shows that, inside the first core, the azimuthal velocity gradually increases with time and has a nearly Keplerian velocity inside the inner disk. 
In addition, Figure~\ref{fig:5}{\it d} shows that the azimuthal velocity is over 10 times larger than the radial velocity inside the first core.
Thus, at these epochs,  the gas orbits around the protostar with a nearly Keplerian velocity inside the first core.

Two nested structures (i.e. the first core and the inner disk) merge to form a single disk $\tc=3.396$\,yr after the protostar formation (green line in Fig.~\ref{fig:5}{\it a}).
The green lines in Figure~\ref{fig:5}{\it c} indicate that the merged disk is supported by rotation and has a size of $\sim1$\,AU.
After the merger, the disk increases its mass and size with time.
This disk corresponds to the circumstellar disk.
At the end of the calculation ($\tc=3.5$\,yr), the circumstellar disk has a mass of $0.025\msun$ with a size of $1.1$\,AU, while the protostar has a mass of $3.5\times10^{-3}\msun$ with a size of 0.01\,AU.
Thus, the circumstellar disk is approximately seven times more massive than the protostar.

\subsection{Evolution of Cloud with Extremely Slow Rotation and No Rotation}
We also calculated the evolution of the circumstellar disk in the molecular cloud with a further lowering of the rotational energy of $\beta_0=10^{-5}$.
However, the formation process of the circumstellar disk is almost the same as that of the model with $\beta_0=10^{-4}$.
Also, in this model, two nested disks appear just after the protostar formation, and finally they merged to form a single circumstellar disk.
The circumstellar disk has a mass of $0.016\msun$ with a size of $0.3$\,AU, while the protostar has a mass of $4.2\times10^{-3}\msun$ with a size of 0.01\,AU at the end of the calculation ($\tc=3.7$\,yr).

On the other hand, when the initial cloud has no angular momentum, only a single core appears outside the protostar.
The time sequence for evolution of the collapsing cloud  after the protostar formation for the model without rotation ($\beta_0=0$, model 4) is shown in Figure~\ref{fig:6}.
The azimuthally averaged distributions of the density and velocity for three different epochs corresponding to each panel in Figure~\ref{fig:6} are plotted in Figure~\ref{fig:7}.
Also, in this model, the first core appears in the collapsing cloud before the protostar formation.
Figure~\ref{fig:6}{\it a} indicates that the first core exists without disappearing at the protostar formation epoch. 
In Figure~\ref{fig:7}, at the protostar formation epoch ($\tc=0$, black line), there are two shock surfaces: the outer shock corresponding to the first core is located at $r\sim1$\,AU, while the inner shock corresponding to the protostar is located at $r\sim0.01$\,AU. 
\citet{masunaga00} showed that the first core falls onto the protostar and disappears for a short duration after the protostar formation.
Figure~\ref{fig:6} indicates that, after the protostar formation, the first core shrinks with time and finally disappears in $\tc\sim1.3$\,yr.
Figure~\ref{fig:7} shows that the shock surface corresponding to the first core moves inwardly and merges with the protostar.
Thus, no circumstellar disk appears for the model without rotation.
We also confirmed that, after the first core disappears, the density distribution in the infalling envelope is proportional to $ \rho \propto r^{-1.5}$ in the range of $0.01\,{\rm AU} < r <10 $\,AU, while it is proportional to $ \rho \propto r^{-2}$ in the range of $r>10$\,AU.
This distribution resembles the singular isothermal self-similar solution \citep{shu77}.

\subsection{Evolution of Cloud with a Relatively Rapid Rotation}
As shown in \S\ref{sec:mod}, when the initial cloud has a considerably slow rotation ($\beta_0=10^{-4}$), two nested disks (the first core and the inner disk) co-exist for a short duration after the protostar formation, in which the inner disk is enclosed by the outer first core.
On the other hand, when the cloud has a rapid rotation, we cannot distinguish the first core and the inner disk in the star formation process.
In other words, only a single rotating disk (or rotating first core) appears before and after the protostar formation.
To investigate the formation of the circumstellar disk for the cloud with a relatively rapid rotation, the cloud evolution for the model with $\beta_0=10^{-3}$ (model 1) is presented in this subsection.

The density and velocity distributions 16.7\,yr (i.e., $\tc=-16.7$\,yr) before the protostar formation for model 1 ($\beta_0=10^{-3}$) are shown in Figure~\ref{fig:8}.
The figure shows that the first core has a size of $\sim8$\,AU and has already become sufficiently thin before the protostar formation.
The velocity vectors in this figure indicate that the gas is rapidly rotating inside the first core.
Figure~\ref{fig:9} shows the density and velocity distribution for model 1 $\tc=1.475$\,yr after the protostar formation with various spatial scales.
The entire region of the first core is covered in Figure~\ref{fig:9}{\it a}, while the protostar embedded in the first core is seen in Figure~\ref{fig:9}{\it d}.
At this epoch ($\tc=1.475$\,yr), the first core and protostar have sizes of $\sim8$\,AU and $\sim0.01$\,AU, respectively.
The arrows in Figures~\ref{fig:9}{\it a}-{\it d} indicate that the gas is rapidly rotating inside the first core outside the protostar.
Although the spiral structure appears inside the first core as seen in each upper panel of Figures~\ref{fig:9}{\it b} and {\it c},  no clear shock corresponding to the inner disk as seen in Figures~\ref{fig:1} and \ref{fig:2} appears in this figure.
As shown in Figure~\ref{fig:9}{\it d}, the gas falls onto the protostar rapidly in the vertical direction and very slowly in the horizontal direction.
In addition, the disk-like first core sags downwards in the proximity of the protostar, as seen in lower panels of Figures~\ref{fig:9}{\it c} and {\it d}.
In Figure~\ref{fig:10}, the density and velocity distributions before (upper panel) and after (lower panel) the protostar formation are plotted in three dimensions.
The figure shows that the first core has a sufficiently thin disk-like structure before the protostar formation.
In addition, just after the protostar formation, the protostar is already enclosed by the thin disk.
In the bottom wall, we can also see the spiral structure that is developed by the excess angular momentum of the disk (or the first core).

In Figure~\ref{fig:11}, the azimuthally averaged density ({\it a}), radial velocity ({\it b}),  azimuthal and Keplerian velocities ({\it c}) and ratio of the radial to the azimuthal velocity ({\it d}) are plotted against the radius for three different epochs [the epoch before ({\it black}) and after ({\it red}) the protostar formation and just before the protostar formation epoch ({\it blue})].
In Figure~\ref{fig:11}{\it a}, the shock at $\sim 8$\,AU corresponds to the surface of the first core and remains almost in the same position by the end of the calculation.
Inside the first core ($\lesssim8$\,AU), the gas density increases smoothly towards the centre of the cloud before the protostar formation (black and blue lines), while it increases with waves after the protostar formation (red line).
This is because the spiral structure develops inside the first core after the protostar formation, as seen in Figure~\ref{fig:9}.
Figure~\ref{fig:11}{\it b} shows that the radial velocity suddenly increases to approach zero at the surface of the first core 
($\sim 8$\,AU) in each epoch.
Just after the first core formation (black line), the radial velocity becomes $v_r\sim0$ inside the first core, indicating that the gas collapses very slowly towards the centre of the cloud.
As the central density increases, the second collapse is induced and the gas collapses rapidly again, as seen in the blue line of Figure~\ref{fig:11}{\it b}.
After the protostar formation, the radial velocity inside the first core oscillates around $v_r = 0$, because the spiral structure appearing after the protostar formation contributes to the angular momentum transfer.
Thus, inside the disk (or the first core), a fraction of gas can fall onto the protostar, while the remaining gas moves  outward with an excess angular momentum.
However, as seen in Figure~\ref{fig:11}{\it d}, because the azimuthal velocity greatly dominates the radial velocity inside the first core, the radial motion is not much noticeable inside the first core.

Figure~\ref{fig:11}{\it c} shows that, in each epoch, the azimuthal velocity traces a near-Keplerian velocity inside the first core.
Thus, the first core already has the Keplerian rotation from its formation.
Then, the first core gradually increases its central density maintaining the Keplerian rotation, and the protostar appears in the Keplerian rotating disk (or the Keplerian rotating first core).
Therefore, as seen in the red line of Figure~\ref{fig:11}{\it c}, the protostar already has a Keplerian rotating disk at its formation.
In other words, the protostar is born in the Keplerian rotating disk that formed long before the protostar formation.
As described in \S\ref{sec:mod}, also in model 2 ($\beta_0 = 10^{-4}$), the protostar is formed inside the disk-like structure (or the disk-like first core).
However, the first core does not reach the Keplerian rotation before the protostar formation (see, Fig.~\ref{fig:5}{\it c}).
Thus, the Keplerian rotating disk (i.e. the inner disk) appears inside the first core after the protostar formation and merges with the first core to form a single Keplerian rotating disk several years after the protostar formation.
This difference is caused by the initial rotational rate of the molecular cloud core.
Because model 1 has a larger rotational energy than model 2, the first core has a Keplerian rotation from its formation and becomes  the circumstellar disk directly after the protostar formation.

At the end of the calculation ($\tc=1.5$\,yr), for model 1, the circumstellar disk has a mass of $0.062\msun$ with a size of $8.5$\,AU, while the protostar has a mass of $2.4\times10^{-3}\msun$ with a size of 0.01\,AU.
Thus, the circumstellar disk is about 25 times more massive than the protostar.
The larger ratio of the mass of circumstellar disk to that of the protostar for this model is also attributed to the larger initial rotation rate.

\section{Discussion}

\subsection{Further Evolution of the Circumstellar Disk}
The formation of the circumstellar disk in the collapsing cloud was investigated in previous studies \citep[e.g.,][]{yorke93,bate98,walch09a,walch09b,machida10}; however, the protostar itself and the region in the proximity of the protostar ($r\ll1$\,AU) were not resolved in such studies.
Thus, the formation and evolution of the circumstellar disk near the protostar just after the protostar formation  have not been understood.
In other words, such studies did not focus on the actual formation process of the circumstellar disk, because the circumstellar disk is smoothly connected to the protostar that has a size of $\sim0.01$\,AU.
In this study, we investigated the formation of the circumstellar disk having Keplerian rotation from the prestellar molecular cloud core stage through the protostar formation resolving the protostar itself, and confirmed that the first core that is formed before the protostar formation is the origin of the circumstellar disk, as expected in \citet{machida10}.
This indicates that  we need to resolve at least the first core to investigate the formation and evolution of the circumstellar disk in the collapsing gas cloud.

When the molecular cloud core has no angular momentum,  the gas inside the first core gradually accretes onto the protostar and the first core disappears in several years after the protostar formation, as shown in \citet{masunaga00}.
On the other hand, the first core does not disappear and evolves into the circumstellar disk after the protostar formation when the molecular cloud core has an angular momentum.

The protostellar mass $M_{\rm ps}$, the mass of the circumstellar disk $M_{\rm disk}$, and circumstellar disk radius $r_{\rm disk}$ at the end of the calculation for each model are listed in Table~\ref{table}.
As noted in the table, the mass of the circumstellar disk is much more massive than that of the protostar by the end of the calculation (or several years after the protostar formation).
The first core that is the origin of the circumstellar disk has a mass of $\sim0.01-0.1\msun$ at its formation, while the protostar has a mass of $\sim10^{-3}\msun$ at its formation. 
The mass of each object (the first core and protostar) is determined by the Jeans mass at its formation.
The Jeans mass continues to decrease as the cloud collapses (see, Fig.2 of \citealt{inutsuka10}), and the first core forms at earlier epoch than the protostar.
Thus, the first core is about $10 - 100$ times more massive than the protostar.
Then, even in the early main accretion phase after the protostar formation, the mass of the circumstellar disk that is directly evolved from the first core is much more massive than that of the protostar.

As shown in \S\ref{sec:mod}, the first core evolves into the circumstellar disk in the main accretion phase even when the molecular cloud core has a considerably smaller angular momentum of $\beta_0\simeq10^{-4}-10^{-5}$ whose value is comparable or smaller than the lower limit of observations $\beta_0 \sim 10^{-4}$ \citep{goodman93,caselli02}.
This indicates that, in general, a massive Keplerian disk rapidly (or suddenly) appears just after (or before) the star formation. 
In the main accretion phase subsequent to the gas-collapsing phase, both the circumstellar disk and the protostar increase their mass by gas accretion.
In this study, we could not investigate further evolution of the circumstellar disk in the main accretion phase, because we calculated its formation and evolution with a sufficiently high-spatial resolution that has a considerably short timestep.
On the other hand, \citet{machida10} calculated the formation and evolution of the circumstellar disk with a relatively coarser spatial resolution at the expense of the structure in the proximity of the protostar, and  showed that the circumstellar disk is more massive or comparable to the protostar by the end of the main accretion phase.
It is expected that such a massive disk tends to show fragmentation and subsequent formation of the binary companions or planet-size objects in the circumstellar disk.
Even when no fragmentation occurs in the main accretion phase, such a massive disk is a favourable site for planet formation after the main accretion phase, because gas-giant planets can be formed by gravitational instability \citep{durisen07}.
However, we require a huge amount of CPU time to investigate the further evolution of the disk after the main accretion phase from the molecular cloud core stage.

\subsection{Fragmentation in the Early Collapsing Phase}
\label{sec:frag}
In this study, to suppress fragmentation and binary formation, we adopted a relatively small initial rotational energy ($\beta_0\le10^{-3}$) that is smaller than the average of the observations $\beta_0\sim0.02$ \citep{goodman93,caselli02}.
Recent numerical simulations have shown that fragmentation or binary formation occurs just after the first core formation when the initial cloud has a relatively larger rotational energy of $\beta_0>10^{-3}$ (\citealt{matsu03}, see also \citealt{bodenheimer00,goodwin07}).
Thus, the mass and size of the fragments are comparable to those of the first core ($r \sim 1-10$\,AU, $M\sim0.1-0.01\msun$), because the first core fragments to form a few clumps.
In addition, although the angular momentum is redistributed into the orbital and spin angular momenta after fragmentation, each fragment still has sufficient angular momentum to form the Keplerian-rotating disk inside it, as shown in \citet{machida05b}.
Therefore, it is expected that the circumstellar disk forms in each fragment passing through the same evolutional path as the non-fragmentation model, and the circumstellar disk in each fragment has a similar size and mass to the non-fragmentation models.
Note that the disk-formation process may differ from non-fragmentation models when fragmentation occurs inside the first core (or in the higher-density region), because the properties of the fragment  are different from those of the first core.
However, also note that, \cite{bate98} showed that fragmentation rarely occurs in such a high-density region ($n\gg10^{10}\cm$) in the collapsing cloud core.

We can expect that, in each fragment, the circumstellar disk with Keplerian rotation appears after the protostar formation, and such a disk is more massive than (or comparable to) the protostar in the early main accretion phase, as shown in \S\ref{sec:results}.
However, as the circumstellar disks grow, the disks in each fragment may interact.
Thus, we do not know whether the circumstellar disk in each fragment traces the same evolutional path for non-fragmentation model in the main accretion and subsequent phases.
To investigate the evolution of the circumstellar disk in the binary or multiple systems, further long-term calculations that cover binary or multiple systems are necessary.

\subsection{Effects of Magnetic Field}
In this study, we adopted unmagnetized cloud cores as the initial state.
However, the magnetic field plays a significant role in the star-formation process.
For example, fragmentation or binary formation is suppressed by the magnetic braking \citep{machida08a,hennebelle08}, and the protostellar outflows are driven by the magnetic effect \citep{machida07,hennebelle2008a}.

In the collapsing cloud core, however, the magnetic field in the high-density gas region ($n\gtrsim10^{12}\cm$) dissipates by the Ohmic dissipation \citep{nakano02}.
Thus, a very weak field exists inside the first core (or the region around the protostar, see \citealt{machida07}).
In this study, we only calculated the evolution of the circumstellar disk for a short duration, in which the circumstellar disk has a density of $n\gtrsim 10^{12}\cm$, as seen in Figures~\ref{fig:1}, \ref{fig:2} and \ref{fig:9}.
Therefore, in the early main accretion phase on which this study focuses, the magnetic field may hardly affect the evolution and formation of the circumstellar disk, because the disk formation occurs in the high-density gas region where the magnetic field is decoupled from neutral gas.

The circumstellar disk increases its size and mass by the end of the main accretion phase as shown in \citet{machida10}, in which the evolution of the circumstellar disk in the unmagnetized cloud was investigated.
However, when the cloud is strongly magnetized, the magnetic field may suppress further evolution of the disk.
Recently, \citet{mellon09} and \citet{duffin09} proposed that a larger size disk exceeding $\sim10$\,AU may not be formed in a strongly magnetized cloud core, because the angular momentum of the infalling gas is effectively transferred by magnetic braking.
As the circumstellar disk grows by gas accretion, the low-density gas region ($n\ll10^{12}\cm$) appears, where the magnetic field can be coupled with neutral gas and the angular momentum is transferred by magnetic braking.
In addition, in the magnetized collapsing cloud, the circumstellar disk (or first core) can drive the protostellar outflow that also outwardly transfers the angular momentum from the parent cloud, and may suppress the larger disk formation \citep{tomisaka02,machida04,machida06b,machida09a}.
However, in the later gas-accretion phase, the gas with larger specific angular momentum falls onto the circumstellar disk, because the outer envelope has a larger angular momentum and the gas of the outer envelope falls onto the centre of the cloud later.
Such infalling gas with larger specific angular momentum may be able to contribute to the larger size disk formation, over the magnetic braking catastrophe.
To investigate the actual size of the circumstellar disk, we need to calculate the disk evolution for the further long term in the magnetized cloud core.

\section{Summary}
In this study, we calculated the evolution of the collapsing gas cloud from the molecular cloud core until the circumstellar disk with Keplerian rotation appears resolving both molecular cloud core and  protostar itself, and found that the circumstellar disk originated from the first core and is much more massive than the protostar in the early main accretion phase.
In summary, the massive Keplerian disk already exists just after the protostar formation.
This result is different from the classical picture of the circumstellar disk formation, in which the circumstellar disk gradually grows its size and mass after the protostar formation.

The formation of the circumstellar disk from the prestellar core stage is summarized in Figure~\ref{fig:12}, in which three different epochs for first core formation, protostar formation and main accretion phases are schematically described for three different modes of the protostar and circumstellar disk formation (i.e. for the cloud with rapid [{\it a}], slow rotation [{\it b}] and no rotation [{\it c}]). 
The figure shows that the first core appears before the protostar formation.
The first core shrinks and disappears in several years after the star formation when the molecular cloud has no angular momentum (Fig.~\ref{fig:12}{\it c}).
On the other hand, when the molecular cloud has an angular momentum, the first core has a disk-like structure at its formation and remains without disappearing even after the protostar formation.

When the cloud has a relatively rapid rotation (Fig.~\ref{fig:12}{\it a}), the first core already has a Keplerian rotation from its formation.
Then, the first core gradually increases its central density, and the protostar appears when the number density exceeds $n\gtrsim10^{20}\cm$.
Thus, at the protostar-formation epoch, the protostar is already surrounded by the Keplerian rotating disk, which is about 100 times more massive than the protostar.
In summary, the Keplerian disk formation precedes the protostar formation.
In other words, the protostar is born inside the Keplerian rotating disk.

\begin{figure}
\includegraphics[width=150mm]{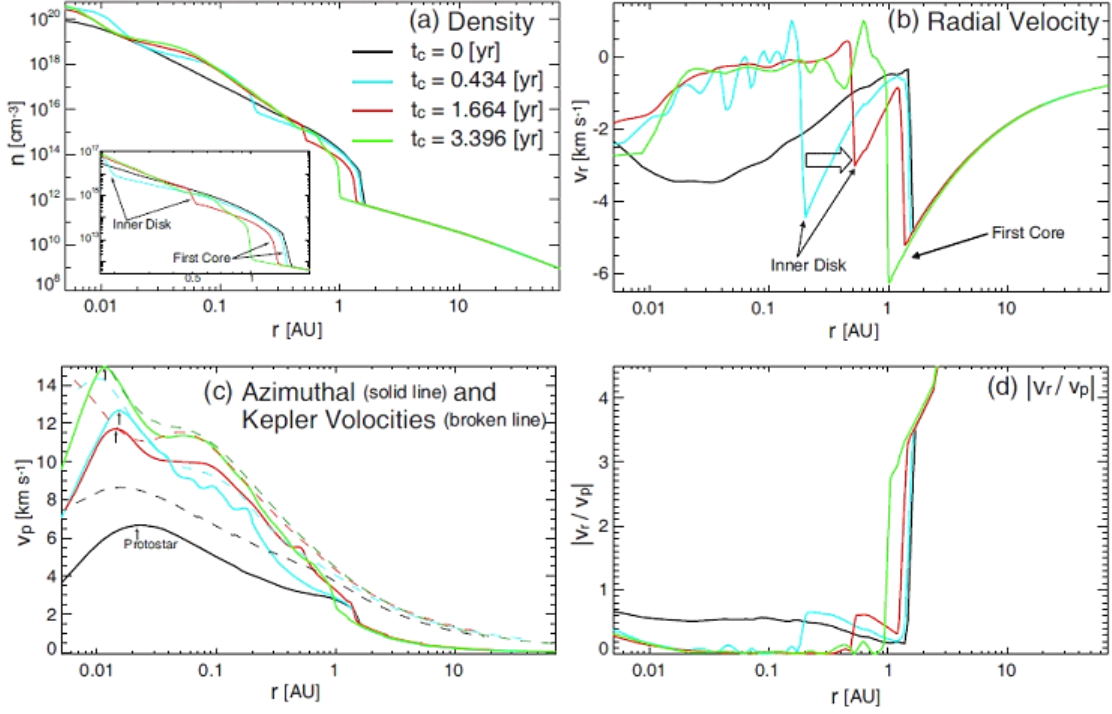}
\caption{
Radial distribution of (a) the density, (b) radial velocity, (c) azimuthal velocity and (d) the ratio of radial velocity to azimuthal velocity for 
model 2 ($\beta_0=10^{-4}$) with different epochs.
The lower-left inset in panel (a) is a close-up view in the range of $0.02\,{\rm AU} < r <  0.2\,{\rm AU}$.
The dotted line in panel (c) is the Kepler velocity at each epoch.
}
\label{fig:5}
\end{figure}

\begin{figure}
\includegraphics[width=150mm]{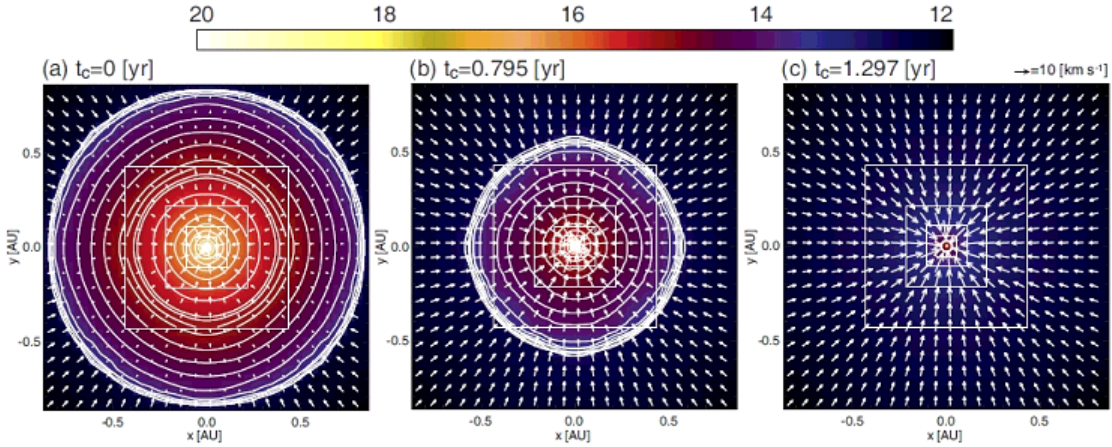}
\caption{
Time sequence for model 4 ($\beta_0=0$) after the protostar formation.
The density distribution ({\it colour and contours}) and velocity vectors ({\it arrows}) on the equatorial plane are plotted in each panel.
The elapsed time $\tc$ after the protostar formation is described in each panel.
}
\label{fig:6}
\end{figure}

\begin{figure}
\includegraphics[width=100mm]{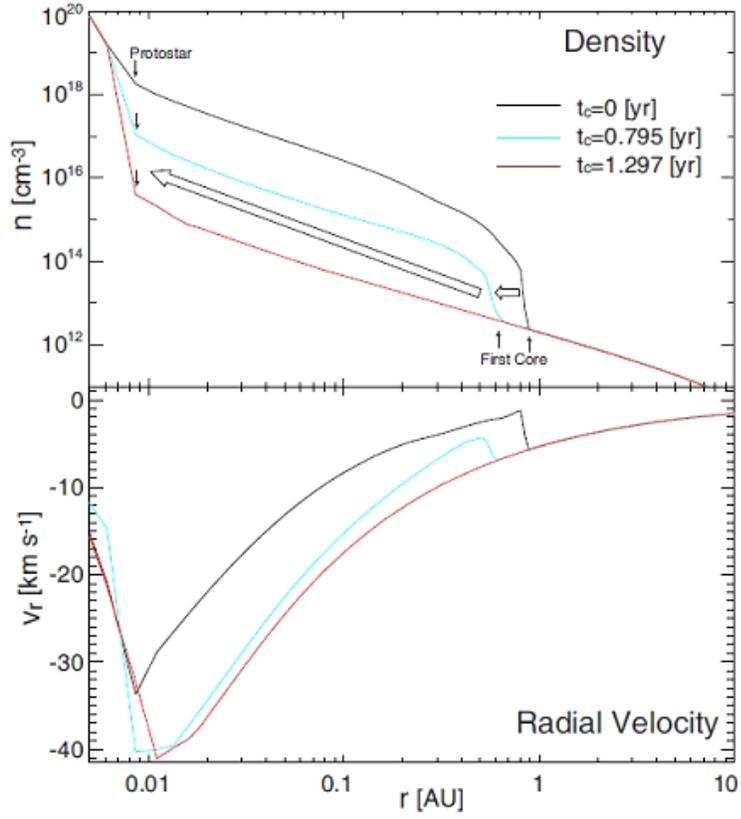}
\caption{
Distribution of density (upper panel) and radial velocity (lower panel) for model 4 ($\beta_0=0$).
}
\label{fig:7}
\end{figure}

\begin{figure}
\includegraphics[width=100mm]{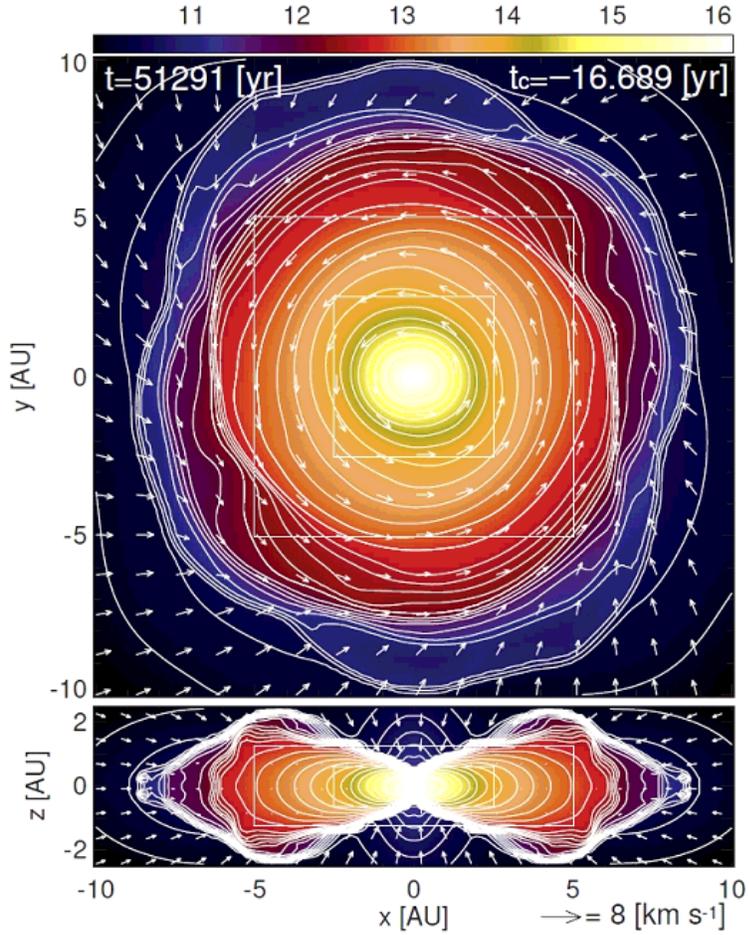}
\caption{
Density distribution ({\it colour and contours}) and velocity vectors ({\it arrows}) on the 
$z=0$ (upper panel) and $y=0$ (lower panel) planes for model 1 ($\beta_0=10^{-3}$) before the protostar formation.
}
\label{fig:8}
\end{figure}

\begin{figure}
\includegraphics[width=150mm]{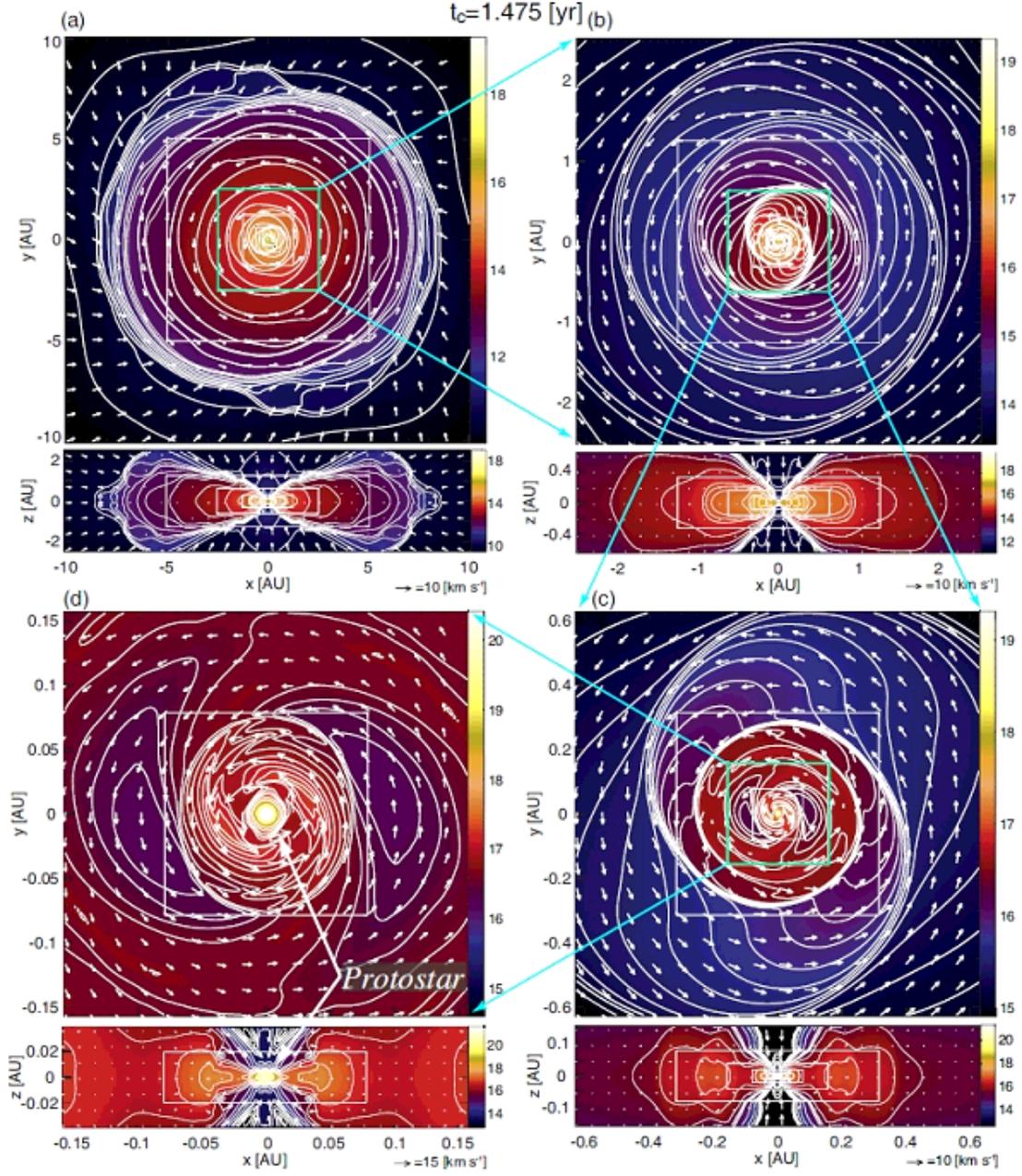}
\caption{
Density distribution ({\it colour and contours}) and velocity vectors ({\it arrows}) on $z=0$ (each upper panel) and $y=0$ (each lower panel) plane at $\tc = 1.474$\,yr for model 1 ($\beta_0=10^{-3}$) with different spatial scales.
}
\label{fig:9}
\end{figure}

\begin{figure}
\includegraphics[width=120mm]{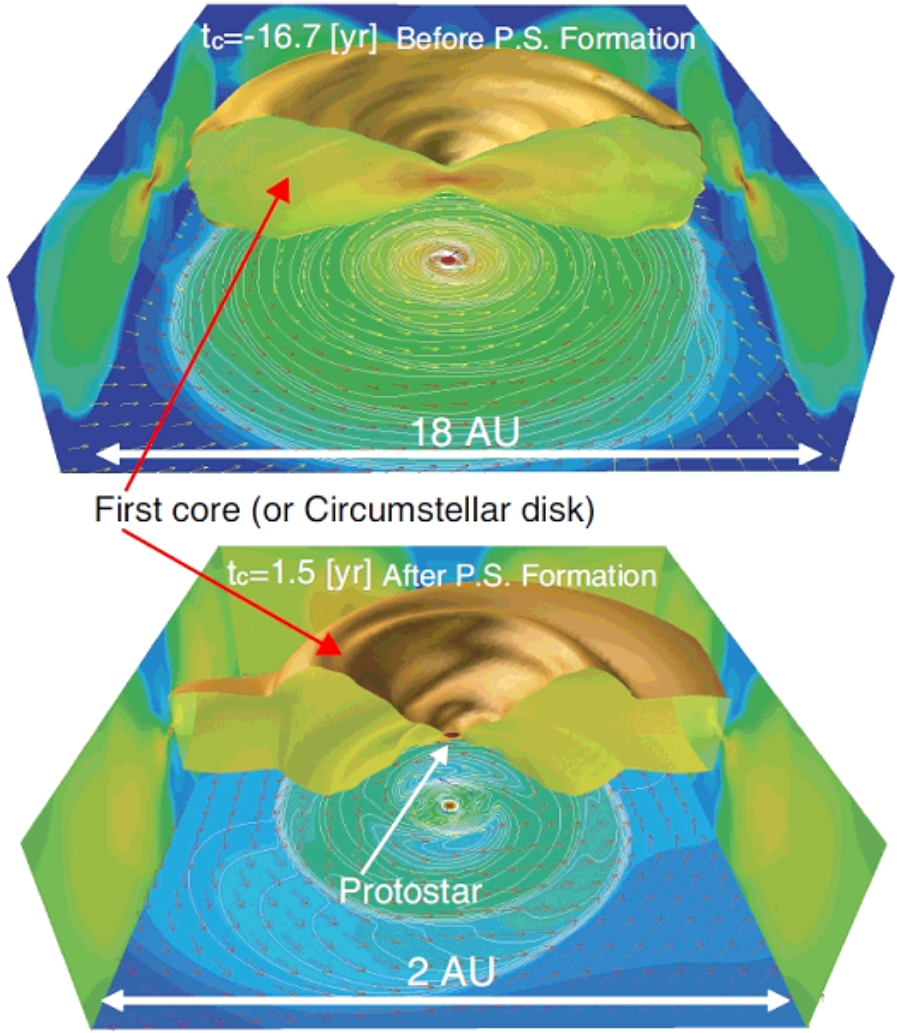}
\caption{
Density distribution ({\it isodensity surface}) around the centre of the cloud before (upper panel) and after (lower panel) the protostar formation.
The density distribution on the $x=0$, $y=0$ and $z=0$ plane is projected onto each wall surface.
The velocity vectors on the $z=0$ plane are also projected onto the bottom wall surface.
}
\label{fig:10}
\end{figure}

\begin{figure}
\includegraphics[width=150mm]{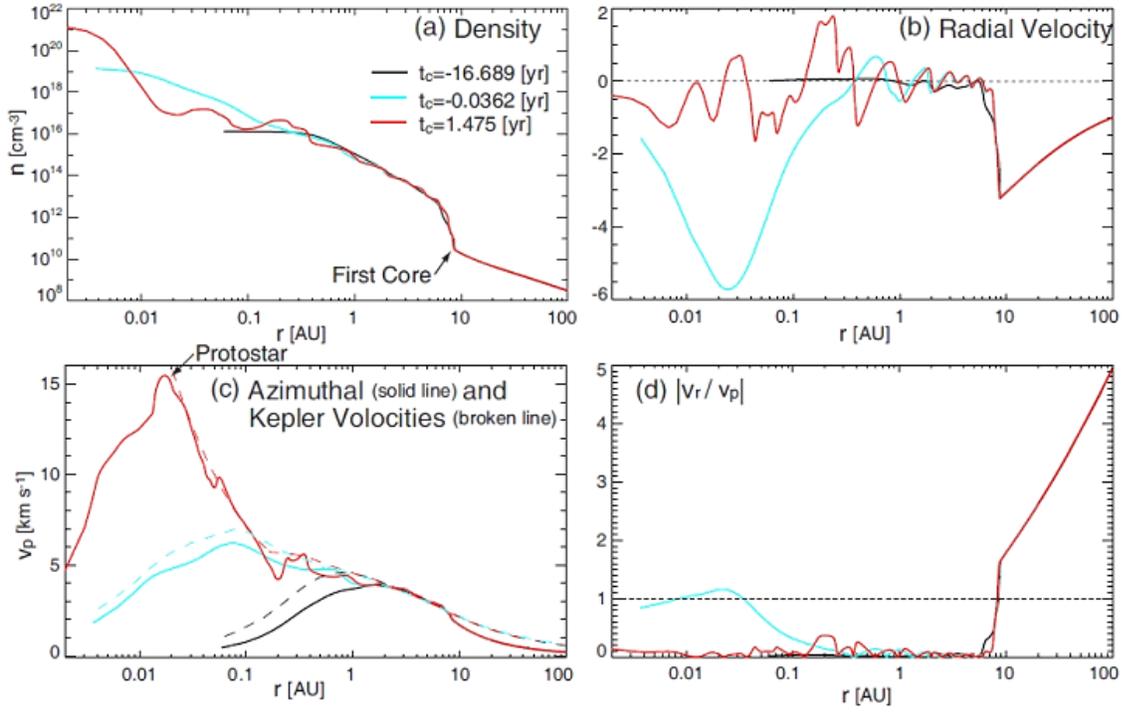}
\caption{
Radial distribution of (a) density, (b) radial velocity, (c) azimuthal velocity and (d) the ratio of radial velocity to azimuthal velocity for 
model 1 ($\beta_0=10^{-3}$) with different epochs.
}
\label{fig:11}
\end{figure}

\begin{figure}
\includegraphics[width=150mm]{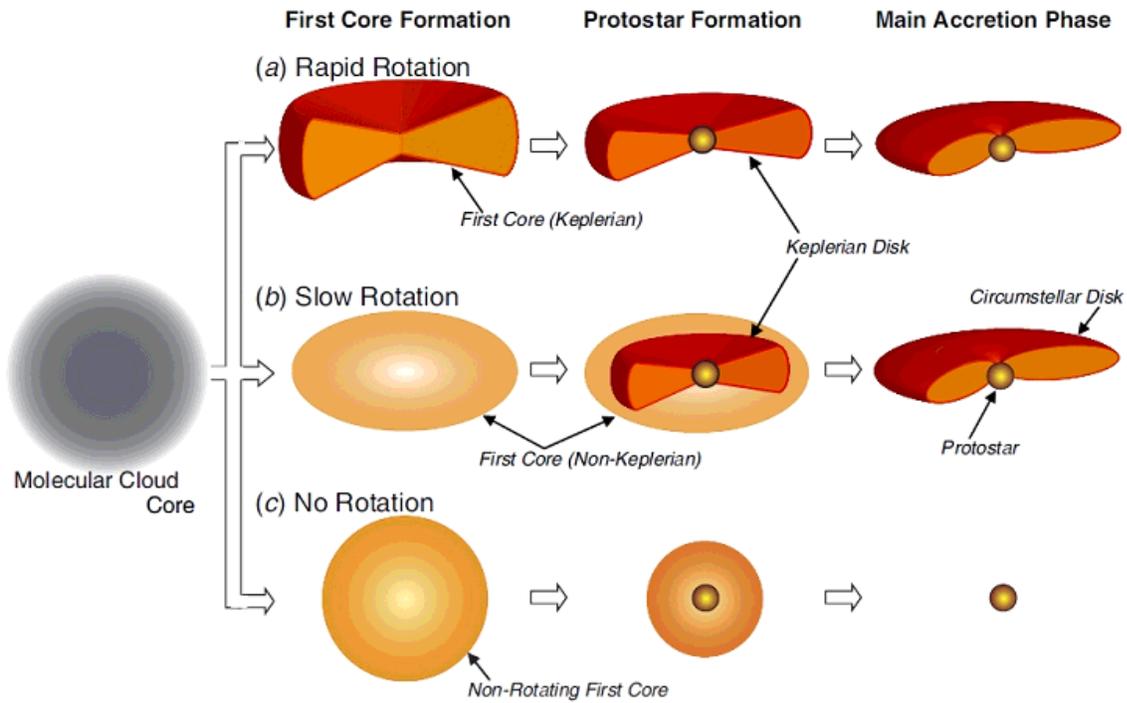}
\caption{
Schematic view of the circumstellar disk formation in the collapsing cloud core, showing three different evolutional sequences for the cloud with (a) rapid rotation, (b) slow rotation, and (c) no rotation.  
}
\label{fig:12}
\end{figure}

When the molecular cloud has a relatively slow rotation rate (Fig.~\ref{fig:12}{\it b}), the first core is partially supported  by the centrifugal force, and mainly by thermal-pressure-gradient force.
Thus, at the first-core-formation epoch, although the first core has an oblate structure, it rotates slowly with a sub-Keplerian velocity.
In this case, after the protostar formation, two nested cores (or two nested disks) appear as seen in Figure~\ref{fig:12}{\it b}.
The outer disk corresponds to the remnant of the first core, while the inner disk corresponds to the Keplerian disk formed in the proximity of the protostar.
In the main accretion phase, the gas of the first core gradually falls onto the protostar.
Because the sizes of the first core ($\sim 1$\,AU) and the protostar ($\sim0.01$\,AU) are considerably different, the infalling gas increases its rotational velocity as it shrinks in the radial direction according to the angular momentum conservation law.
As a result, the Keplerian disk (i.e. the inner disk) appears around the protostar.
Then, the inner disk increases its size, while the first core gradually shrinks.
Finally, two nested disks merge to form a single disk that has a Keplerian velocity and corresponds to the circumstellar disk.
At the epoch of the merger, the size and mass of the merged disk are almost the same as those of first core at its formation.
Note that the first core shrinks before merger, while it still has a size of $\sim1$\,AU (Fig.\ref{fig:5}). 
Thus, also in a cloud with considerably small rotational energy, the massive Keplerian disk already exists just after the protostar formation.

In this study, we showed that a massive disk already exists before the protostar formation, and  the protostar is enclosed by the massive Keplerian circumstellar disk from the moment of its birth.
Although we only calculated the evolution of the circumstellar disk several years after the protostar formation, we can expect that the circumstellar disk increases its size by gas accretion, as in the classical picture of circumstellar disk formation.
In addition, our result indicates that the massive disk ($>0.1-0.01\msun$) with a size of $>1$\,AU already exists in the very early phase of the star formation 
(or before the star formation).
We expect that such a disk can be detected using future instruments such as ALMA.

\acknowledgments
We have greatly benefited from discussions with ~K. Omukai.
Numerical computations were carried out on NEC SX-9 at Center for Computational Astrophysics, CfCA, of National Astronomical Observatory of Japan, and NEC SX-8 at the Yukawa Institute Computer Facility.
This work was supported by the Grants-in-Aid from MEXT (20540238, 21740136).

\begin{table}
\setlength{\tabcolsep}{3pt}
\caption{Model parameters and calculation results}
\label{table}
\footnotesize
\begin{center}
\begin{tabular}{c|cc|cccccccccc} \hline
{\footnotesize Model} & $\beta_0$ &  $\Omega_0$ {\scriptsize [s$^{-1}$]}& $M_{\rm ps}$ {\scriptsize [$\msun$]} 
& $M_{\rm disk}$ {\scriptsize [$\msun$]}  & $r_{\rm disk}$ {\tiny [AU]} \\ \hline
1     & $10^{-3}$ & $6.5 \times 10^{-14}$ & $2.4\times10^{-3}$ & 0.062 & 8.5 \\
2     & $10^{-4}$ & $2.1 \times 10^{-14}$ & $3.5\times10^{-3}$ & 0.025 & 1.1 \\
3     & $10^{-5}$ & $6.5 \times 10^{-15}$ & $4.2\times10^{-3}$ & 0.016 & 0.3 \\
4     & 0 &  0 & $3.4\times10^{-3}$  & 0 & 0 \\
\hline
\end{tabular}
\end{center}
\end{table}

\clearpage

\end{document}